\documentstyle[moriond,12pt]{article}
\begin{document}
\titlepage
\heading{CLUSTERING STATISTICS \& DYNAMICS}

\author{R. Juszkiewicz$^{1}$, F. R. Bouchet$^{2}$}
       {$^{1}$ Copernicus Center, Warsaw, Poland.}
       {$^{2}$ Institut d'Astrophysique de Paris, France.}
\begin{abstract}{\baselineskip 0.4cm 
Since the appearance of the classical paper of
Lifshitz almost half a century ago,
linear stability analysis of 
cosmological models is textbook knowledge. Until recently, however,
little was known about the behavior of higher than linear order terms
in the perturbative expansion. These terms become important in the
weakly nonlinear regime of gravitational clustering, when the rms
mass density contrast is only slightly smaller than unity.
In the past, theorists showed little interest in studying
this regime, and for a good reason:
only a decade ago, it would have been an academic excercise --
at scales large enough to probe the weakly nonlinear regime, all
measures of clustering were dominated by noise.
This is no longer the case with present data.
The purpose of this talk is to provide a brief summary
of recent advances in weakly nonlinear
perturbation theory. 
We present analytical perturbative results 
together with results of 
N-body experiments, conducted to test
their accuracy. We compare perturbative predictions
with measurements from galaxy surveys. Such comparisons
can be used to test the gravitational instability theory and to constrain
possible deviations from Gaussian statistics
in the initial mass distribution; they can be also used to study the
nature of physical processes that govern galaxy formation 
(``biasing''). We also show how 
future studies of velocity field statistics can
provide a new way to determine the density parameter, $\Omega$.
}
\end{abstract}

\bigskip
\bigskip
\bigskip
\bigskip
\bigskip
\bigskip
\bigskip
\bigskip
\bigskip
\bigskip
\bigskip

\centerline{To appear in: {\em Clustering in the Universe},}
\centerline{Proc. XXX Moriond Meeting, Les Arcs, 1995,}
\centerline{Eds. S. Maurogordato {\em et al.}} 
\centerline{Editions Frontieres, Paris}

\eject

\section{Perturbation theory}

Tests of theories for the origin of the large scale structure of the
universe ultimately depend on a comparison of model predictions with 
measurable quantities, derived from observations.
The statistical measures we will discuss here are the low order
N-point correlation functions. They have two clear advantages.
First, they can be estimated from galaxy surveys
with a reasonable degree of precision and reproducibility
(see, e.g. \cite{{lss1},{lss2},{obs2},{obs3}},
and references therein). Second, correlation functions can be 
relatively easily related to dynamics (e.g. \cite{lss1}). Their evolution
can be studied by taking moments of the hydrodynamical equations
of motion of an expanding self-gravitating pressureless fluid
with zero vorticity, which  is a good 
approximation of the real universe after the hydrogen recombination.
Low order correlation functions can also be measured from N-body
experiments. 

\subsection{Gravitational instability}

The full description of gravitational instability is nonlinear. The
density contrast
\begin{equation}
\delta({\bf x},t) \; = \; {\rho({\bf x},t)\over\langle\rho\rangle}
- 1 \; ,
\label{eq:def}
\end{equation}
the peculiar velocity ${\bf v}$ and the gravitational potential $\phi$
are related by the Euler, Poisson and continuity equations. They
can be combined into one expression for the density contrast,
\begin{equation}
{\textstyle
\ddot\delta + 2H\dot\delta
- {3\over2}\Omega H^2 \delta \; = \;
{3\over2}\Omega H^2 \delta^2 + a^{-2}\nabla\delta\cdot\nabla\phi
+ a^{-2}\nabla_{\alpha} \nabla_{\beta} [(1 + \delta)v_{\alpha} 
v_{\beta}] \; .
}
\label{eq:motion}
\end{equation}
Here ${\bf x} = \{ x_{\alpha} \}$ and $t$ are,
respectively, the comoving spatial coordinates and the cosmological
time, the dots represent time derivatives,
$\; \nabla_{\alpha} = \partial /\partial x_{\alpha} \;$,
$a(t)$ is the scale factor, $H =\dot a/a$ is the Hubble parameter,
and $\Omega = 8\pi G \langle \rho \rangle / 3H^2$ is the density
parameter.
The brackets $\langle\ldots\rangle$ denote ensemble averaging and
$\rho$ is the mass density. Perturbation theory rests on a
conjecture that when the deviations from homogeneity are small,
the first few terms of the expansion 
\begin{equation}
\delta \; = \; \delta_1 + \delta_2 + \delta_3 + \ldots 
\label{eq:series}
\end{equation}
provide a reasonable approximation of the exact solution
of eq. (\ref{eq:motion}).
The first, linear term is the well known Lifshitz \cite{el}
solution of eq. (\ref{eq:motion}) with the right-hand side set to zero,
\begin{equation}
\delta_1 \; = \; D(t)\varepsilon({\bf x}) \; + \;\;{\rm decaying \; mode} \; ,
\label{eq:linear}
\end{equation}
where $D(t)$ is the standard growing mode (see, e.g., \S 11 in \cite{lss1})
and $\varepsilon$ is a
random field with statistical properties defined by initial conditions.
The term $\, \delta_2 = {\cal O}
(\varepsilon^2) \, $ is the solution of the same equation with
quadratic nonlinearities included iteratively by using $\delta_1$
as source terms (as in \cite{lss1}, \S 18). For a vanishing cosmological
constant ($\Lambda = 0$) and arbitrary $\Omega$, the fastest
growing mode in this solution is
\cite{j3}
\begin{equation}
{\textstyle
\delta_2 \; = \; D^2(t)\,\left[{2\over3}(1+\kappa)\varepsilon^2 +
\nabla\varepsilon\cdot\nabla\Phi + ({1\over2} - \kappa)
\tau_{\alpha\beta}\tau_{\alpha\beta}\right] \; ,
}
\label{eq:quad}
\end{equation}
\begin{equation}
\;\;\; \textstyle{
\tau_{\alpha\beta} \; = \;({1\over3}\delta_{\alpha\beta}\nabla^2
-\nabla_{\alpha}\nabla_{\beta})\Phi \; ; \;\;\;
\Phi({\bf x}) \; = \;  - \, \int\,d^3x'\,\varepsilon({\bf x'})/4\pi
|{\bf x - x'}| \; .
}
\label{eq:tau}
\end{equation}
The parameter $\kappa $ is a slowly varying
function of time; its dependence on $\Omega$ is extremely weak,
too. In the range $0.05 \leq \Omega \leq
3$, the $\Omega$-dependence is well approximated by \cite{frb2}
\begin{equation}
\textstyle
\kappa\left[ \Omega(t) \right] \; \approx \; {3\over14}\,\Omega^{-2/63}
\; .
\label{eq:kappa}
\end{equation}
The higher order terms can be constructed out of the linear solution
in a similar way. 

\subsection{Dynamical evolution of correlations}

The main source of difficulties in studies of
the dynamics of gravitational instability is
the nonlinearity of the equations of motion. 
Indeed, let us write eq. (\ref{eq:motion}) for $\delta({\bf x}, t)$
and $\delta' = \delta({\bf x'}, t)$, then multiply
the first equation by $\delta'$, and the second by $\delta$.
The result of adding these two equations and averaging is
\begin{equation}
\ddot\xi + 2H\dot\xi - 3\Omega H^2 \xi
+ 2 a^{-2}\nabla^2 \xi_v \; = \; 3\Omega H^2 \langle\delta^2\delta'\rangle
\; + \; \ldots \; ,
\label{eq:master}
\end{equation}
where $\, \xi = \langle\delta\delta'\rangle \, $ and
$\, \xi_v = \langle {\bf v \cdot v'}\rangle \, $ are the
two-point density and velocity correlation functions, while
the right-hand side of the equation involves 3- and 4-point
correlations, which we left out except for one term given
as an example. The full expression is not the point here;
the important message is that we have an infinite hierarchy
of equations. Mathematically, this is similar to the BBGKY 
hierarchy: in order to solve for an $N$-point function,
we need to know $(N+1)$- and $(N+2)$-point correlations.
Perturbation theory may help, but not without further
assumptions, necessary to close the infinite hierarchy
of correlation functions. Here we will assume
that
$\varepsilon({\bf x})$ is a random Gaussian field.
This approach, also known as the
{\it random phase approximation}, can be justified theoretically for
a wide class of models of the early universe
\cite{lss2}. Independently of whether we
take ``inflationary'' arguments seriously or not,
the random phase assumption is observationally
testable. As we will show below, comparing
model predictions with observations can help us
decide whether the early universe was Gaussian or not.
The greatest advantage of our assumption is simplicity. Indeed: 
all odd-order correlation functions of $\varepsilon$ vanish, and all
even-order moments can be reduced to the two-point
function. For example, for the first four moments we have
$
\langle \, 1 \,\rangle \, = \, \langle \, 123 \, \rangle \, = \, 
0$, $\, \langle \, 12 \,\rangle 
\, = \, \xi_L\,(|{\bf x}_1 - {\bf x}_2|,t) \, D^{-2} \,$ , 
and
\begin{equation}
\langle \, 1234  \, \rangle = D^{-4}\,\xi_L \, (|{\bf x}_1 -
{\bf x}_2|,t)
\, \xi_L \, (|{\bf x}_3 - {\bf x}_4|,t) \;\;
+ \;\; {\rm cycl.\,(two \; terms)} \; ,
\label{eq:even}
\end{equation}
where $\, \langle \, 12 \ldots N \, \rangle
\, \equiv \, \langle \, \varepsilon({\bf x}_1)\varepsilon({\bf x}_2)
\ldots \varepsilon({\bf x}_N) \, \rangle$,
while $\xi_L$ in eq. (\ref{eq:even}) is $\, \langle \, 
\delta_1 \, ({\bf x}_1,t) \,
\delta_1 \, ({\bf x}_2,t) \, \rangle$, the linear 2-point
correlation function.
Using these relations and the perturbative expansion, we can
express the 3- and 4-point functions on the right-hand side of 
eq. (\ref{eq:master}) in terms of $\xi_L$. This will close the
hierarchy, allowing us
to calculate weakly nonlinear corrections for the 2-point function
$\xi$. The same procedure can be applied for higher order 
correlation functions. This simple idea is at the root of all
perturbative results, discussed below.

Perturbative calculations are usually easier after applying
the Fourier transform. The Fourier transform of $\xi$ is the
power spectrum, $P(k,t)$, where $k$ is the comoving wavenumber. 
The linear term in the perturbative
expansion for $P(k,t)$ is 
\begin{equation}
P_L\,(k,t) \; = \;
\int \, \xi_L\, ({\bf r},t) \, \exp(i{\bf k \cdot r}) 
\, {\rm d}^3k \; . 
\label{eq:power}
\end{equation}
The basic properties of linear evolution,
as well as the effect of nonlinear interactions can be
deduced from the functional form of $\delta_1$ and $\delta_2$.
Indeed, according to eq. (\ref{eq:linear}), all fluctuations
evolve without changing the shape of their density profiles,
so that we can expect that the linear power spectrum evolves without
changing its slope (the growth rate is the same for
all scales). This is indeed the case. If $P_L\,(k,t) \propto k^n$
at some initial time $t_0$, than at later times $P_L\,(k,t) =
P_L\,(k,t_0)\,D^2(t)/D^2(t_0) \, \propto k^n$. A second characteristic
property of a linear field is that its distribution function
remains Gaussian; only its width, $\langle \delta_1^2\rangle $
grows in time as $D^2(t)$. Neither of these properties
remain valid when nonlinear interactions become important.
The nonlinear terms, describing tidal interactions in eq.
(\ref{eq:quad}) redistribute power across the
spectrum, and growth rates may be enhanced at some wavenumbers
and suppressed at others: the logarithmic slope $n(k,t) = {\rm d}(\log \,
P)/{\rm d}(\log k)$ is no longer constant, it can now change with time.
Moreover, the nonlinear evolution introduces deviations
from Gaussian statistics. The lowest-order moment which can
detect such deviations is the skewness, $\langle\delta^3\rangle$.
The Gaussian distribution is symmetric about $\delta = 0$,
and its skewness is zero. According to eq. (\ref{eq:quad}), 
however, the fluctuations with positive initial density
contrast $\delta_1$ grow more rapidly than those with
$\delta_1 < 0$. Hence, nonlinear interactions introduce
asymmetry in the distribution, and we can expect that
with time, gravity will induce a non-zero skewness in
the distribution. 

The earliest (to our knowledge) analytic results
of weakly nonlinear calculations were obtained by
Zeldovich et al. and Peebles \cite{{zel1},{dor},{lss1}}.
They considered power-law initial conditions, $P_L \, \propto
\, k^n$, and showed that 
unless $n < 4$, the nonlinear terms become larger than
linear at large scales, 
and the perturbation theory breaks down. 
A somewhat more systematic approach, going beyond scaling
relations was introduced in the eighties by
Juszkiewicz et al. \cite{{j1},{j2}} and Vishniac 
\cite{vis}, who used Eulerian perturbation theory and
the random phase approximation to study the first non-trivial,
weakly-nonlinear corrections for second moments -- $
P(k,t)$ and $\xi(r,t)$. Peebles \cite{lss1} used the same
approach to calculate the gravitationally
induced skewness, while Fry, in a truly seminal paper, 
showed how to calculate
a reduced correlation function of arbitrary order $N$, using
the perturbative expansion, truncated at the $(N - 1)$ -st
order \cite{jf1}.

Unfortunately, N-body simulations, available in the early eighties were
too crude to study the range of validity of perturbative methods. 
Moreover, the galaxy surveys at scales, large enough to probe
the weakly nonlinear regime of clustering, were dominated by noise.
This made further progress in analytic calculations
difficult and the field became dormant
for several years. With time, the dynamical range 
of the simulations improved by more than an order of magnitude 
\cite{nbody2}. Simultaneously,
the length scales in galaxy surveys, at which correlations can
be accurately measured, increased by 
an order of magnitude \cite{obs2}, 
making weakly nonlinear theory useful in their analysis.
As a result of these developments, the study of the evolution
of the mass distribution in the expanding universe has 
become a rich topic for analytic perturbative methods.
The revival of enthusiasm for analytic calculations produced 
many new results. It is our purpose to briefly summarize 
some of these new results here. For a more detailed approach, see e.g.
the recent lecture notes of Bouchet \cite{varenna} and the review by Sahni \&
Coles \cite{varun}.  

\subsection{Spatial smoothing}

The last issue we need to discuss before moving on to the
next section is spatial smoothing. It is useful to employ
a field $\bar\delta$, such that $\delta({\bf x},t)$ is
replaced by $\bar\delta_1 + \bar\delta_2 + \ldots$, with
\begin{equation}
\bar\delta_N ({\bf x},t) \; = \; \int \, \delta_N({\bf x'},t)\,w(|{\bf
x - x'}|\,/R)\,{\rm d}^3x' \; ,
\label{eq:filter}
\end{equation}
where $w(x/R)$ is a spherically symmetric window function.
$R$ is the characteristic comoving smoothing length: $w$ decreases
rapidly for $x/R \gg 1$. Its Fourier representation
acts as a low pass filter, suppressing the fuctuations
with wavenumbers $k \gg 1/R$. The usual normalization
condition requires $\int \, {\rm d}^3r \,w(r) = 1$.
Smoothing over a ball of radius $R$ is also called  ``top hat''
smoothing; another popular filter is a fuzzy ball with
a Gaussian profile of width $R$. Note that the effects
of smoothing and gravitational evolution are interchangeable
only for $\delta_1$; for terms of second or higher order,
these two effects {\em do not} commute (this is evident from
eq. (\ref{eq:quad}); see also \cite{gr1}).  For a power-law  initial
spectrum, the smoothed linear variance is
\begin{equation}
\langle\bar\delta^2_1\rangle \, = \, \sigma_L^2(R,t) \, = \,
\int w(x_1/R)w(x_2/R)\xi_L(|{\bf x}_1 - {\bf x}_2|)
\,{\rm d}^3x_1{\rm d}^3x_2 \; \propto \, R^{-(n+3)}
\; ,
\label{eq:sigmaL}
\end{equation} 
so the condition $n > -3$ ensures that the clustering process
is hierarchical, and proceeds from small to large scales.
(this constraint, together with the Zeldovich-Peebles condition 
$n < 4$, broadly determines the range of ``reasonable'' values of $n$;
a mix of observational arguments and theoretical
prejudce shrinks this range to $-2 \lsim n \lsim 1$, see
e.g. \cite{lss2}). The smoothing procedure
is needed for at least two reasons. First, at any stage
of the evolution, in order to be able to use the perturbation
theory, we need to smooth-out the strong nonlinearities.
In all hierarchical models, this can be done by any
a low pass filter of width $R > R_{nl}$, where $R_{nl}(t)$ is the
current nonlinear scale, defined by the condition
$\sigma^2_L(R_{nl},t) = 1$. The second reason for introducing the
smoothing is practical: 
all observable quantities, derived from galaxy surveys or
numerical experiments, involve spatial smoothing.
Hereafter, we will deal with smoothed fields almost
without exceptions (which will be explicitly stated),
so to simplify notation, we will omit the bars over $\delta$
and $\delta_N$. Generally, we will also assume 
Gaussian initial conditions 
(again, all exceptions will be properly described).

\section{New results for the second moment; the previrialization conjecture}

The perturbative series (\ref{eq:series}) can be used to expand the
second moment of the density field in powers of the
linear variance, $\sigma^2_L$.
Squaring $\delta$ and averaging, we get
\begin{equation} 
\sigma^2 \; \equiv \; \langle \delta^{2} \rangle 
\; = \; \langle \delta_{1}^{2} \rangle + 
\langle \delta_{2}^{2} \rangle + 2 \langle
\delta_{1} \delta_{3} \rangle + \ldots
\; = \; \sigma_L^2 + (I_{22} + I_{13})\, \sigma_L^4
+ {\cal O}(\sigma_L^{6}).  
\label{eq:Iij}
\end{equation}
The second and third term in this expansion
are the lowest order nonlinear corrections to linear theory.
These terms are of order $\sigma_L^4$ because
${\cal O}(\sigma_L^3)$ terms vanish under the random phase
assumption. 
The quantities $I_{22}$ and $I_{13}$ can be determined from
perturbation theory. They were recently calculated for
power-law initial conditions by Scoccimarro \& Frieman \cite{sco}
and by {\L}okas et al. \cite{l2}. Scoccimarro \& Frieman
explored the evolution of an unsmoothed density field. {\L}okas et al.
studied spatially smoothed fields, and
found that nonlinear tidal interactions redistribute
the power in the spectrum in such a way, that for
$n > -1$, the correction $I_{22} + I_{13}$ is negative.
The growth rate of density fluctuations is inhibited:
$\sigma$ is smaller than the linear prediction $\sigma_L$.
The magnitude of this suppression increases with $n$,
or with the relative amount of small scale power in the
initial spectrum. For $n < -1$, there is an opposite
effect -- the fluctuation growth rate is enhanced
$(\sigma > \sigma_L)$. The transition occurs at 
$n = -1$, when the correction is close to zero and
$\sigma \approx \sigma_L$. {\L}okas et al. compared
these perturbative results with a series of N-body simulations
and found a good agreement in a wide dynamical range, up
to $\sigma \approx 1$. These results are in
qualitative agreement with the so-called {\em previrialization
conjecture} of Davis \& Peebles \cite{dav}. According
to Davis \& Peebles, tidally induced nonradial motions of small scale
inhomogeneities within collapsing
mass concentrations should resist gravity and slow
down the collapse. Until recently the size of
the previrialization effect was still under discussion,
with some N-body studies reproducing the effect and
others claiming its absence. For example, Peebles 
\cite{pjep} found strong support for the previrialization
effect, while Evrard \& Crone \cite{evr} reached opposite
conclusions. This apparent controversy can be resolved,
using the perturbative results we just discussed.
Indeed, Peebles in his simulations assumed $n = 0$,
while Evrard \& Crone assumed $n = -1$: a spectral index,
for which nonlinear effects are minimal. 

Jain \& Bertschinger \cite{jb} used perturbation theory
and N-body experiments to investigate the
nonlinear transfer of power in a CDM spectrum. Their 
results are consistent with {\L}okas et al. and
Scoccimarro \& Frieman. Finally, we must mention
the important work of Makino et al. \cite{mak},
who succeeded in reducing the dimensionality of the
mode coupling integrals of 
Vishniac and Juszkiewicz et al. \cite{{vis},{j2}}. 
This significantly simplified the derivation of all
of the new results described in this section.

\section{The third moment and deviations from Gaussian behavior}

As we saw in Section 1, nonlinear gravitational evolution
drives the field $\delta$ away from the initial Gaussian
distribution. Deviations from Gaussianity can be measured
by {\it cumulants}, also called ``reduced'' or ``connected''
moments. A cumulant of order $N$ is
\begin{equation}
M_N \; = \; \partial^N\ln\,\langle e^{x\delta}
\rangle /\partial x^N \; , \qquad {\rm evaluated
\;\; at} \;\; x = 0 \; .
\label{eq:cumulant}
\end{equation}
For the first four cumulants, this expression generates
the mean, $M_1 = \langle\delta\rangle = 0$, the variance,
$M_2 = \langle \delta^2 \rangle = \sigma^2$, the skewness, 
$M_3 = \langle \delta^3 \rangle$, and the kurtosis,
$M_4 = \langle \delta^4 \rangle - 3\sigma^4$. For a zero-mean
Gaussian distribution, all cumulants vanish except $M_2$.
The skewness measures the asymmetry of the distribution,
while the kurtosis measures the flattening of the tails relative
to a Gaussian. Using perturbation theory, 
Fry \cite{jf1} showed that in an Einstein-de Sitter universe,
$M_N$ scale like $\sigma^{2N - 2}$. Hence, 
to lowest non-vanishing order in perturbation theory, the ratio
\begin{equation}
S_N \; = \; M_N\,/\,\sigma^{2N - 2} 
\label{eq:sl}
\end{equation}
is time-independent. Juszkiewicz et al. \cite{j4}
have recently used this property and 
the Edgeworth series to describe the weakly nonlinear
evolution of the probability distribution function
(PDF) of the density field. The Edgeworth expansion is
a standard statistical tool, used to represent
quasi-gaussian probability distribution functions (PDFs).
Juszkiewicz et al. obtained an expansion in powers 
$\sigma$, with the first few terms,
given by 
\begin{equation}
{\textstyle
p(\nu) \; = \; \left[ \, 1 + {1\over 3!}\,S_3 H_3(\nu)\sigma
+ \left\{{1\over 4!}\,S_4H_4(\nu) + {10\over 6!}\,S_3^2H_6(\nu)\right\}
\sigma^2 \,\right]\,g(\nu) \, + \, {\cal O}\left(g(\nu)\sigma^4\right)
\; ,
\label{eq:edge}
}
\end{equation}
where $\nu = \delta/\sigma$, $g = (2\pi)^{-1/2}\exp(-\nu^2/2)$,
and $H_J$ is a Hermite polynomial of degree $J$.
A similar expansion was obtained independently by Bernardeau \&
Kofman \cite{fb3}. The Edgeworth expansion is particuraly useful
when only a few low-order reduced moments are known; it
can be also used to estimate low-order $S_N$ -s from the shape
of empirically measured PDF near the peak (in realistic situations,
the tails, and/or high-$N$ moments are poorly known anyway).
An alternative, Laplace-transform 
representation for the PDF, which does require
the knowledge of all $S_N$ -s to $N = \infty$, was proposed
by Bernardeau \cite{fb4}.

\subsection{Skewness induced by gravity}

The acual values of the parameters $S_N$ can be also calculated
perturbatively. The results generally depend on the shape of
the initial power spectrum and the shape of the window function.
Let us begin with the skewness, responsible for 
the lowest order nonlinear correction to the Gaussian PDF in
eq. (\ref{eq:edge}).
Upon raising both sides of equation (\ref{eq:series}) to the third power
and averaging, we get 
\begin{equation}
\langle \delta^3 \rangle \; = \; S_3\, \sigma^4 \;
= \; \langle\delta_1^3\rangle \, + \, \langle 3\delta_1^2\delta_2 \rangle
\, + \,\ldots
\label{eq:skew}
\end{equation}
In models, which satisfy the random phase assumption, the lowest
order non-vanishing term in this expansion is ${\cal O}(\sigma)^4$.
The scaling with $\sigma$ is completely different in models with 
non-gaussian initial conditions and an
intrinsic skewness $\langle\delta_1^3\rangle \ne 0$:
the lowest order term is $\propto \sigma^3$.  The skewness parameter,
$S_3$ in eq. (\ref{eq:skew}) was first calculated by Peebles
\cite{lss1}, assuming $\Omega = 1$ and no smoothing. The latter
constraint prevents any comparison of this result with observations
or N-body simulations. To deal with this problem, we calculated
$S_3$ for a smoothed density field with a power-law initial spectrum
and for two different kinds of spatial filters: a gaussian and a
top hat \cite{j3}.
The result is particularly simple in the latter case,
\begin{equation}
{\textstyle
S_3 \;
 = \; {34\over 7} \, - \, (n+3) \, + \, {\cal O}(\sigma^2)
\; ,
}
\label{eq:S3}
\end{equation}
where the allowed range
of power indexes is $-3 \leq n < 1$ (outside of this range
$S_3$ diverges). Note that here, as well as in Section 2,
we quote only results for the Einstein-de Sitter case. 
All these results, however,  are also true for a 
much wider class of models with
arbitrary $\Omega$ and $\Lambda$
(this is an approximate statement, but the approximation
is good enough for all practical purposes). 
For example, for $\;0 \leq \Lambda/3H^2 \leq 1 - \Omega\;$ 
and $\;0.05 \leq \Omega \leq
3\;$, the term `34/7' in eq. (\ref{eq:S3}) should be
replaced by 
\begin{equation}
{\textstyle
4 + 4\kappa(\Omega) \; = \;
{34\over 7} \; + \;{6\over 7}\,(\Omega^{-0.03} - 1)
\; .}
\label{eq:curvature}
\end{equation}
The corrections, introduced by deviations from the Einstein-de Sitter
model were first derived by Bouchet et al. \cite{frb2} for
arbitrary $\Omega$, and by Bernardeau \cite{fb1} and
Bouchet et al. for arbitrary $\Lambda$ \cite{frb4}.
These results were recently confirmed by 
Catelan et al. \cite{ca1}, who used a significantly different
method of derivation. For our purposes here, the most important
implication of all this is that Gaussian initial conditions
and weakly nonlinear
gravitational instability generate the scaling relation
\begin{equation}
M_N \; \propto \; \sigma^{2N-2} \qquad\qquad (\sigma < 1) \; ,
\label{eq:scaling}
\end{equation}
and $S_N$, the ratio of the above moments, up to terms of order
$\sigma^2$ in perturbation theory is time-, $\Omega$- and
$\Lambda$-independent; it is determined only
by the shape of the initial power spectrum.
Bernardeau \cite{fb1} generalized eq. (\ref{eq:S3}) 
for a smoothing scale-dependent slope, $n = n(R)$.
He showed, that this expression for $S_3(R)$ remains valid
when $n$ is replaced by the effective local slope,
\begin{equation}
n(R) \; = \; - \,{\rm d}(\log \sigma^2_L(R))/
{\rm d}(\log R) - 3 \; .
\label{eq:nR}
\end{equation}
As we have seen, to lowest non-vanishing
order in perturbation theory, $S_3$ is time independent. We
can expect that $S_3(R)$ will be scale-independent as well in the
range of smoothing scales, for which $n(R)$ 
is constant. A completely different picture arises
when the primeval PDF is significantly skew. The lowest order
non-vanishing contribution to skewness gives 
\begin{equation}
S_3(R) \; \propto \; 1/\sigma_L \; \propto \;
R^{-(n+3)/2}D(t)^{-1} \; , 
\label{eq:blowup}
\end{equation}
so in all hierarchical
($n > -3$) models we can expect a strong scale- and time-dependence,
with $S_3$ ``blowing up'' at large scales and at high redshift.
As pointed out 
by several authors, this property can be used to distinguish
observationally
Gaussian initial conditions from their non-gaussian alternatives
\cite{{co2},{j3},{sil}}. Chodorowski \&
Bouchet \cite{chodor} find that the distinction can be made even sharper
by studying the scaling properties of $S_4$ (see also
Fry \& Scherrer \cite{jf3}). 

Using methods similar to those
discussed above, Catelan \& Moscardini \cite{mosca}, {\L}okas
et al. \cite{l1}, and Bernardeau \cite{fb1} calculated
the parameter $S_4$ for power-law spectra and top hat as well as
Gaussian windows. The last two papers
also provide closed-form expressions for $S_4 = S_4(n)$.
Bernardeau \cite{fb4} solved the problem of calculating
$S_N$ -s of arbitrarily high order 
for top hat smoothing.

Most of the above perturbative results for $S_3$ and $S_4$
were tested against N-body simulations 
\cite{{ba},{frb1},{j3},{j4},{ol},{l1},{wein}}, 
showing remarkable agreement up to $\sigma \approx 1$.
On a comoving scale $R$,
the Edgeworth expansion agrees with the N-body simulations
within $\sim 1/\sigma(R)$ standard deviations from zero
\cite{j4}. The Laplace transform perturbative PDF agrees with
the N-body PDF even when $\sigma$ is slightly greater
than unity \cite{fb4}.

\subsection{Velocity statistics} 

Unlike the reduced moments
of the density field, the cumulants of the {\it expansion
scalar}, $\theta = \nabla\cdot{\bf v}/H$ are very sensitive
to the value of $\Omega$. For example, the skewness parameter
is given by the expression \cite{fb2}
\begin{equation}
{\textstyle
\langle\theta^3\rangle \,/\, \langle\theta^2\rangle^2 \;
= \; - \,\Omega^{-0.6}\,\left[\,{26\over 7} -
(n + 3)\,\right] \; .}
\label{eq:scalar}
\end{equation}
In principle, this can provide a new method of determining
$\Omega$ from the velocity data alone (rather than from the
comparison with the density field, as in more conventional
methods). The existing velocity surveys, however, are by far
too shallow to allow accurate measurements of the variance and
skewness of $\theta$. We have to wait for better
data. Weakly nonlinear theory can also be used to study
the dynamics of relative motions in pairs of galaxies,
which induce the anisotropy of the redshift space 
galaxy correlations \cite{j5}. This can improve
the accuracy of $\Omega$ determinations from 
presently available redshift surveys.

\subsection{Lagrangian methods}

So far, we have focused
on Eulerian perturbation theory. An interesting alternative
is the Lagrangian approach. It follows particle trajectories
rather than the evolution of $\delta({\bf x}, t)$. 
The local density is $\delta = 1/J - 1$, where $J$
is the Jacobian of the transformation from initial
particle positions to positions at time $t$.
The famous Zeldovich approximation \cite{zel2} is a linear
order Lagrange solution. In its original form, it
can not be used for statistical studies because it
predicts infinite cumulants for $\delta$. 
As pointed out by Grinstein \& Wise \cite{gr2}, the problem can
be at least partially fixed by expanding $1/J - 1$
in powers of spatial derivatives of the linear order
Eulerian velocity field. The cumulants,
generated by such an expansion are finite.
Unfortunately, however, they disagree with the proper
Eulerian perturbation theory and with all N-body experiments
\cite{{mosca},{j3},{j4}}. The Zeldovich approximation
fails particularly badly when applied to the velocity divergence,
$\theta \;$ \cite{{fb2},{j4}}. Two lessons follow from this example.
The first is that dynamical estimates of $\Omega$, derived from 
reconstruction schemes for the initial PDF of $\delta$
or $\theta$, which rely entirely on the Zeldovich approximation
(see e.g. \cite{avi}), should
be treated with caution. The second lesson is positive:
the Zeldovich approach is more powerful than it seems. 
For example, under this approximation, $\kappa = 0$ in eq.
(\ref{eq:curvature}) and therefore for a top hat
$S_3 = 4 - (n + 3)$. Depending on $n$, this causes about
20\% to 50\% underestimate of the true $S_3$, derived from
second order Eulerian theory. Note however, that we used {\em first
order} Lagrangian theory to derive the Zeldovich prediction for
$S_3$. To make a fair comparison, we should confront this result
with its first order
Eulerian counterpart: the linear prediction $S_3 = 0$, which is
much further away from the true value! 
What we have just seen, is an illustration of
a more general phenomenon: at any
fixed order in perturbation theory, the Lagrangian
approach is likely to remain valid for stronger
density contrasts than the
Eulerian approach. This happens 
because the requirement of the smallness
of the (Lagrangian) particle displacements is less restrictive
than the requirement of the smallness of $|\delta|$.
This idea, which must have motivated Zeldovich, remains
valid at higher orders in perturbation theory.
The idea to extend the Lagrangian perturbation theory
beyond the linear order was introduced by Moutarde et al. \cite{mou}
and generalized by Bouchet et al. \cite{frb2} and Buchert
\& Ehlers \cite{bt2}. 
Lagrangian coordinates have been used to derive second-
and third-order perturbative solutions in the Einstein-de
Sitter case  \cite{{bt1},{bt2}}, as well as in
more general cosmological models \cite{{frb2},{frb4}},
and to study distortions of correlations
in redshift space \cite{{frb4},{hiv}}. We are
convinced that the full potential of this 
approach to perturbation theory is far from being exhausted.

\subsection{Convergence studies}

As we have seen, perturbative expansions for 
$\langle\delta^2\rangle$
and $\langle\delta^3\rangle$, truncated at the lowest-order
or next to the lowest-order non-vanishing contribution,
are in remarkable agreement with N-body simulations in
a wide dynamical range, up to $\sigma \approx 1$. The precise
range of validity of perturbation theory is still under
investigation. Apart from comparisons with N-body
experiments, the convergence can be also studied within
the perturbation theory itself, by comparing contributions
coming from terms of increasing order. Such studies may
answer the question: why does perturbation theory
work so well?

Scoccimarro \& Frieman \cite{sco} obtained closed-form
expressions for $I_2 = I_{22} + I_{13}$
in the case of no smoothing ($R = 0$).
Although individually, $I_{22}$ and $I_{13}$ suffer from
``infrared'' ($k \rightarrow 0)$ divergencies,
their sum is finite and approximately $n$-independent;
for $\,-2 \leq n \leq 2\,$, the correction term is \cite{{l2},{sco}}
\begin{equation}
I_2 \; \approx \; 1.82 \; .
\end{equation}
The divergent terms in $I_{22}$ and $I_{13}$
cancel each other. The probable source of
this phenomenon is the Galilean invariance of the equations
of motion \cite{sco}. 

In order to make a quantitative
comparison with N-body results possible, 
{\L}okas et al. \cite{l1} included the effects
of smoothing in their calculations. Introduction
of a finite smoothing length ($R \ne 0$)
generates  ``ultraviolet'' ($k \rightarrow \infty$) 
divergences at $n > -1$;
$I_2$ diverges unless $P_L(k)$ has a shortwave cutoff at 
$k = k_c$. For $k_c R \gg 1$,
the correction term is \cite{l2} 
\begin{equation}
I_2 \; \approx \; - (0.6k_c R)^{n+1} \; .
\label{eq:conv}
\end{equation}
We can expect that perturbation theory breaks down when $\sigma_L$
reaches the value, at which the first two terms in the expansion
(\ref{eq:Iij}) become comparable, $|I_2|\sigma_L^4 = \sigma_L^2$.
Therefore, equation (\ref{eq:conv}) suggests that for a given
$k_cR$, the maximal value of $\sigma_L$, determining the
range of validity of perturbation theory, should 
gradually decrease with increasing $n$. 
Comparisons with N-body experiments show that this is
indeed the case \cite{l1}. Note that
a shortwave cutoff naturally appears in N-body experiments; $k_c^{-1}$ 
corresponds to the mean interparticle separation,
or the Nyquist scale.

Scoccimarro \& Frieman \cite{sco} have also calculated
the ${\cal O}(\sigma^2)$ term in the expansion for $S_3$
for the unsmoothed density field. They found that ${\cal O}(\sigma_L^2)$
terms in the expansions for $S_3$ do not ``overtake''
the lowest-order non-vanishing terms until $\sigma_L \approx 0.5$,
in qualitative agreement with the results of direct N-body tests
of perturbation theory, discussed above. A quantitative comparison,
however, will be possible only after smoothing effects are included
(Scoccimarro \& Frieman, in preparation).

\section{Observations}

Before comparing theoretical predictions to observations, let us
discuss two possible sources of confusion.

\bigskip
{\bf 1. Biasing.} If galaxies form with different efficiency in different
environments, then the galaxy distribution may paint a biased picture
of the underlying mass distribution. Let us suppose that the smoothed
galaxy field $\delta_g$ is a local, but not necessarily linear function
of the smoothed mass density field: $\delta_g({\bf x}) =
f[\delta({\bf x})]$, where $f$ is an arbitrary function. Then, using
a Taylor expansion for $|\delta| \ll 1$, it is possible to prove,
as Fry \& Gazta{\~n}aga \cite{jf2} did, that local bias preserves
the form of the scaling relations (\ref{eq:scaling}), while
changing the constants $S_N$. For example, $S_3$ changes to
\cite{{jf2,j4}}
\begin{equation}
S_{3g} \; \equiv \; 
\langle\delta^3_g\rangle\,/\,\langle\delta^2_2\rangle
\; = \; S_3/b_1 \; + \; 3b_2/b_1^2 \; ,
\label{eq:s3g}
\end{equation}
where $b_J \equiv {\rm d}^J f/{\rm d}\delta^J$, evaluated at
$\delta = 0$  (note that the linear bias model with
all $b_J = 0$ for $J > 1$ must break down when $b > 1$
and $\delta < - 1/b$, since it predicts negative $\delta_g$).

\bigskip
{\bf 2. Redshift distortions.} Redshift surveys use radial velocities
of galaxies instead of their true radial coordinates. As a result,
galaxy peculiar motions distort the true spatial distribution and
the two-point correlation function, $\xi$ (cf. \S 20 in 
\cite{lss2} and references therein). Bouchet et al. \cite{frb4}
and Hivon et al. \cite{hiv} have used weakly nonlinear
perturbation theory and N-body simulations
to study the effects of redshift space distortions
on the three-point correlation function, $\zeta$, and on $S_3$.
They showed that although $\zeta$, like $\xi$, is affected
by the redshift space distortion, all appreciable, $\Omega$-dependent
effects cancel out for the moment ratio $S_3$. In the weakly-nonlinear
domain, when $\sigma < 1$, and for spectral slopes
$\, -3 \leq n \leq -1$, the differences between redshift-space $S_3$ 
and  its true value are smaller than 10\%, whatever the value of $\Omega$.
Hence, the comparison between theoretical predictions and observations
is not obscured by redshift space distortions provided we use
the moment ratio $S_3$ rather than reduced moments themselves.

\bigskip
The parameters $S_3$ and $S_4$
have been estimated from the CfA and SSRS surveys
\cite{eg1}, the {\em IRAS} 1.2 Jy sample \cite{frb3}, as well as
for vaious angular catalogues \cite{{eg2},{szap}}. All these 
measurements show beautiful agreement with the predicted scaling
relation (\ref{eq:scaling}).
The dependence (\ref{eq:s3g}) of $S_{3g}$ on the shape of the bias
function may explain why the value measured by Bouchet et al.
\cite{frb3} from the top-hat smoothed, {\em IRAS} redshift survey
is rather low, $\, S^{IRAS}_{3g} \sim 1.5 \,$. The 
galaxies observable in the infrared 
are underrepresented in rich clusters relative to
optical galaxies, so the ``bias function'' that applies to
{\em IRAS} galaxies may have negative curvature ($\, b_2 < 0 \,$),
pushing $S_{3g}$ below the value of $S_3$ for the mass.
If cluster avoidance is indeed the cause of the low value
of $S^{IRAS}_{3g}$, then one would expect that optically selected
surveys would yield higher values. The values of $S_{3g}$,
obtained recently by Gazta{\~n}aga \cite{eg2} from the APM survey
suggest that this is exactly the case. The value of $S_{3g}$,
deprojected from the two-dimensional APM data onto three-dimensional
space is $\, 3.16 \pm 0.14 \,$, assuming a top-hat filter
with a radius $R > 7 \,h^{-1}$Mpc (here $h$ is the Hubble constant
in units $\, 100\,{\rm km\,s^{-1}Mpc^{-1}}$; the latter condition ensures
$\, \sigma_g < 1 \,$). Substituting $n = -1$, appropriate for the
slope, seen in the APM survey on these scales, into equation
(\ref{eq:S3}) yields $\, S_3 = 2.9 \,$. 
Hence, for sufficiently large scales, the APM results are in
perfect agreement with second order perturbation theory:
$\, S_{3g}^{APM} = S_3 \,$. This agreement provides
support for three theoretical assumptions: first, that
gravitational instability is responsible for the growth of
clustering; second, that the initial conditions were Gausssian;
and third, that the APM galaxies trace the mass, i.e.,
in eq. (\ref{eq:s3g}) $\, b_1^{APM} \approx 1 \,$ and
$\, b_2^{APM} \approx 0 \,$ (of course, other combinations
of $b_1$ and $b_2$ also fit the APM data, but they require a coincidental
compensation of the linear, $1/b_1$ factor by the
$b_2$ term arising from nonlinear bias). This is bad news
non-local biasing models like the model of Bower et al.
\cite{syf}, invoked in a desperate attempt to explain
the mismatch of the observed power spectrum with standard
CDM. Indeed, as demonstrated by Frieman \& Gazta{\~n}aga
\cite{frie}, such models break the scaling relation
(\ref{eq:scaling}) and are incompatible with observations.
Most recently, Gazta{\~n}aga \& M{\"a}h{\"o}nen
\cite{eg3} have used the APM data   to
severely constrain the so-called global texture model.
The source of problems for this model is
its prediction that $S_J$
parameters must `blow up' at large scales, like
$S_3$ in our example in eq. (\ref{eq:blowup}).
There is no trace of such behaviour in the APM galaxy
distribution. Since the already available data can be
used to constrain non-local biasing schemes and deviations
from Gaussian distribution, we can expect 
that in the new generations of catalogues
(like the SDSS and the 2dF, described in these proceedings)
will finally tell us whether all we are looking at is just a
Gaussian random process modified by plain gravity.

\acknowledgements{We are grateful to the Organizers,
Chantal Balkowski, Sophie Maurogordato, and
J. Tr{\^ a}n Thanh V{\^ a}n, for making this
scientifically useful meeting happen; to Mother
Nature for providing good skiing conditions, and
last but not least, to Chantal and Sophie as Editors,
for their incredible patience with the authors, who broke several
submission deadlines for this manuscript. This research was supported
by the Polish Government KBN grants No. 2-P304-01607, 2-P304-01707,  
and by an NSF grant No. PHY94-07194. RJ thanks Alain Omont for his
hospitality at Institut d'Astrophysique de Paris, CNRS, where part of the
work reviewed here was done.}

\vfill
\end{document}